\documentclass[3p,times,preprint,twocolumn]{elsarticle}
\usepackage{hyperref}

\usepackage{amsmath,amssymb}
\usepackage{graphicx,tabularx}
\usepackage{epsf}       
\usepackage{epsfig}       
\usepackage[ngerman, english]{babel}
\usepackage{placeins}
\usepackage[
  mode=match,
  propagate-math-font=false,
  reset-math-version=false,
  reset-text-family=true,
  reset-text-series=false,
  text-family-to-math=true,
  text-series-to-math=true]{siunitx}
\usepackage{upgreek}
\usepackage{setspace}
\usepackage{bold-extra}
\usepackage{comment}
\usepackage{multicol}
\usepackage{float}
\usepackage{afterpage}
\usepackage{eurosym}
\usepackage[hang]{footmisc}
\usepackage{lipsum}
\usepackage{subcaption}
\usepackage{color}

\journal{Nuclear Instruments and Methods in Physics Research A}

\begin{document}

\begin{frontmatter}

\title{Characterization of Passive CMOS Strip Detectors After Proton Irradiation}

%\author{A. N. Other \fnref{myfootnote}}
%\address{Physikalisches Institut, Albert-Ludwigs-Universität Freiburg, Hermann-Herder-Straße 3, Freiburg}
%\fntext[myfootnote]{is a cool guy.}


\author[tudortmund]{Marta Baselga\corref{mycorrespondingauthor}}
\cortext[mycorrespondingauthor]{Corresponding author}
\author[desy]{Jan-Hendrik Arling}
\author[desy]{Naomi Davis}
\author[bonn]{Jochen Dingfelder}
\author[desy,bonn]{Ingrid Maria Gregor}
\author[freiburg]{Marc Hauser}
\author[bonn]{Fabian Hügging}
\author[freiburg]{Karl Jakobs}
\author[fhdortmund]{Michael Karagounis}
\author[freiburg]{Roland Koppenhöfer}
\author[tudortmund]{Kevin Alexander Kr\"oninger}
\author[freiburg]{Fabian Lex}
\author[freiburg]{Ulrich Parzefall}
\author[desy]{Simon Spannagel}
\author[freiburg]{Dennis Sperlich}
\author[tudortmund]{Jens Weingarten}
\author[freiburg]{Iveta Zatocilova}
%\author[lastaddress]{Tianyang Wang}

%%%J.-H. Arling, N. Davis, J. Dingfelder, I.-M. Gregor,M. Hauser, F. Hügging, K. Jakobs,  M. Karagounis, R. Koppenhöfer, K. Kröninger, F. Lex, U. Parzefall, S. Spannagel, D. Sperlich, J. Weingarten, I. Zatocilova

\address[tudortmund]{Faculty of Physics, TU Dortmund, Otto-Hahn-Strasse 4a, 44227 Dortmund, Germany} 
\address[desy]{Deutsches Elektronen Synchrotron DESY, Notkestr. 85, Hamburg, Germany}
\address[bonn]{Physikalisches Institut, University of Bonn, Nussallee 12, 53115 Bonn, Germany}
\address[freiburg]{Physikalisches Institut, Albert-Ludwigs-Universität Freiburg, Hermann-Herder-Straße 3, Freiburg, Germany}
\address[fhdortmund]{Fachhochschule Dortmund, Sonnenstraße 96, 44139 Dortmund, Germany} 
%\address[lastaddress]{Zhangjiang Laboratory, No. 99 Haike Road, Zhangjiang Hi-tech Park, Pudong, Shanghai, P.R.China}
%
\begin{abstract}

Strip detectors are populating outer trackers of high-energy particle experiments. They are convenient for covering large areas of sensitive material since they use less power and have fewer readout channels compared to pixels sensors. Nevertheless, they are typically manufactured with a mask set that covers the full wafer, otherwise when using smaller reticles the strip implants have to be stitched. 
%not manufactured by CMOS production lines since they have to be stitched along the implant of the strip and use several reticles to be connected together. 
For this project, strip detectors were fabricated in a CMOS commercial foundry using different reticles to be stitched several times, proving the feasibility of this technology.

LFoundry produced the passive CMOS strip detector with a production line of \SI{150}{\nano\m} node technology, using a \SI{150}{\micro\m} thick FZ wafer. Those strip sensors have three different geometries to study different impacts of the CMOS technology. The strips have lengths of \SI{2.1}{\centi\meter} and \SI{4.1}{\centi\m}, stitching 3 or 5 reticles respectively. This work shows results of \SI{24}{\giga\electronvolt} proton irradiated passive CMOS strip detectors. The detectors were irradiated at CERN and were tested with different set-ups, not showing any effect from the strips stitching.

Proving that this technology is feasible for detecting high-energy particles opens the door to future large productions of passive strip detectors and also to produce active strip sensors in commercial CMOS foundries.

%% previous paper

%Recent advances in CMOS imaging sensor technology , e.g. in CMOS pixel sensors, have proven that the CMOS process is radiation tolerant enough to cope with certain radiation levels required for tracking layers in hadron collider experiments. With the ever-increasing area covered by silicon tracking detectors cost effective alternatives to the current silicon sensors and more integrated designs are desirable.  
%This article describes results obtained from laboratory measurements of silicon strip sensors produced in a passive p-CMOS process. Electrical characterization and charge collection measurements with a $^{90}$Sr source and a laser with infrared wavelength showed no effect of the stitching process on the performance of the sensor.  

\end{abstract}


\end{frontmatter}

%\linenumbers

\section{Introduction}

Future experiments tracking systems for high-energy particle detection will most probably be populated with silicon sensors, since it is advantageous for its availability in semiconductor electronics foundries, the technology of silicon segmentation have been developed during years and it is radiation hard. Detector designs for future accelerators (such as Future Circular Collider FCC, Electron Ion Collider EIC or the Circular Electron Positron Collider CEPC) include tracking layers with silicon detectors. Besides high-energy particle physics experiments, cancer treatment facilities such as hadron therapy centres can use large area strip detectors for beam monitoring and imaging, to provide in time information of the particles while treatment is done. 

Strip detectors are well-suited to cover large area, due to their lower power consumption and fabrication cost with respect to pixel detectors. To date, strip sensors are fabricated in production lines with a mask that covers the full area of the wafer with a photolithography resolution of $\sim$\SI{800}{\nano\m}. % a partir d'aqui és una caca
Using CMOS (Compatible Metal Oxide Semiconductor) production lines for strip sensors could open the door to more processing foundries, with the caveat that strip implants have to be stitched, using multiple reticles for a strip production. 
One of the challenges for passive CMOS strip sensors is proving that the stitching does not affect the performance of the strips. 

\begin{figure}[ht]
    \centering
    \includegraphics[width=\linewidth]{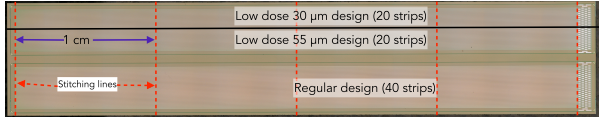}
    \caption{ \small\SI{4.1}{\cm} long strip sensor. It shows the stitching lines and the two different strip designs configuration. 
    \label{fig:long_picture}}
\end{figure}

For this project, two \SI{1}{\cm^2} reticle masks were used to fabricate \SI{2.1}{\cm} and \SI{4.1}{\cm} long strip detectors. Figure \ref{fig:long_picture} shows a picture of a \SI{4.1}{\cm} long strip sensor with the position of the stitching marked with vertical dashed lines. The reticle is repeated four times along the strip. The second reticle has the strip AC and DC pads to one side and the ending of the strip to the other side (the strip end and closing bias and guard rings). 

The strips were fabricated in an LFoundry \cite{lfoundry} \SI{150}{\nano\m} process, stitching the two reticles with a stepper motor. The fabrication used a \SI{150}{\micro\m} thick  p-type float zone wafer with a resistivity around $3-5~\mathrm{k}\Omega\mathrm{cm}$. 

\begin{figure}[ht]
    \centering
    \includegraphics[width=0.95\linewidth]{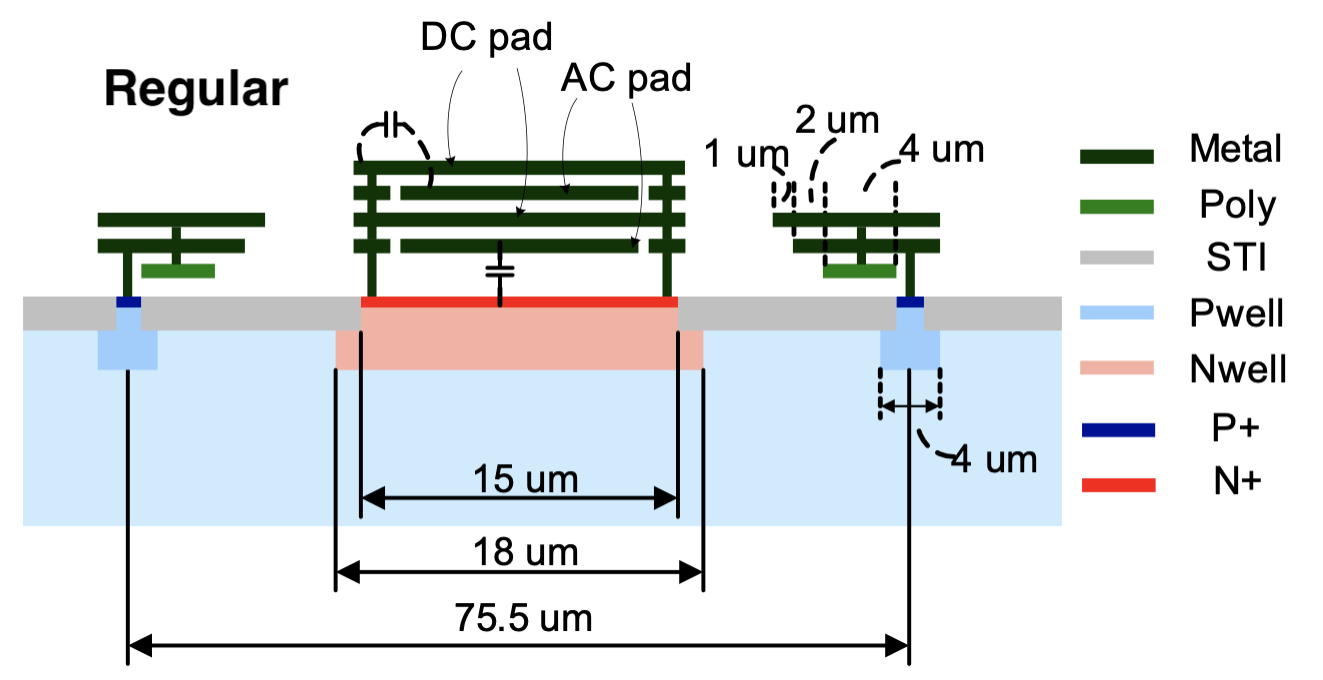}
    \caption{ \small Regular strip design sketch. 
    \label{fig:regular}}
\end{figure}

\begin{figure}[ht]
    \centering
    \includegraphics[width=0.95\linewidth]{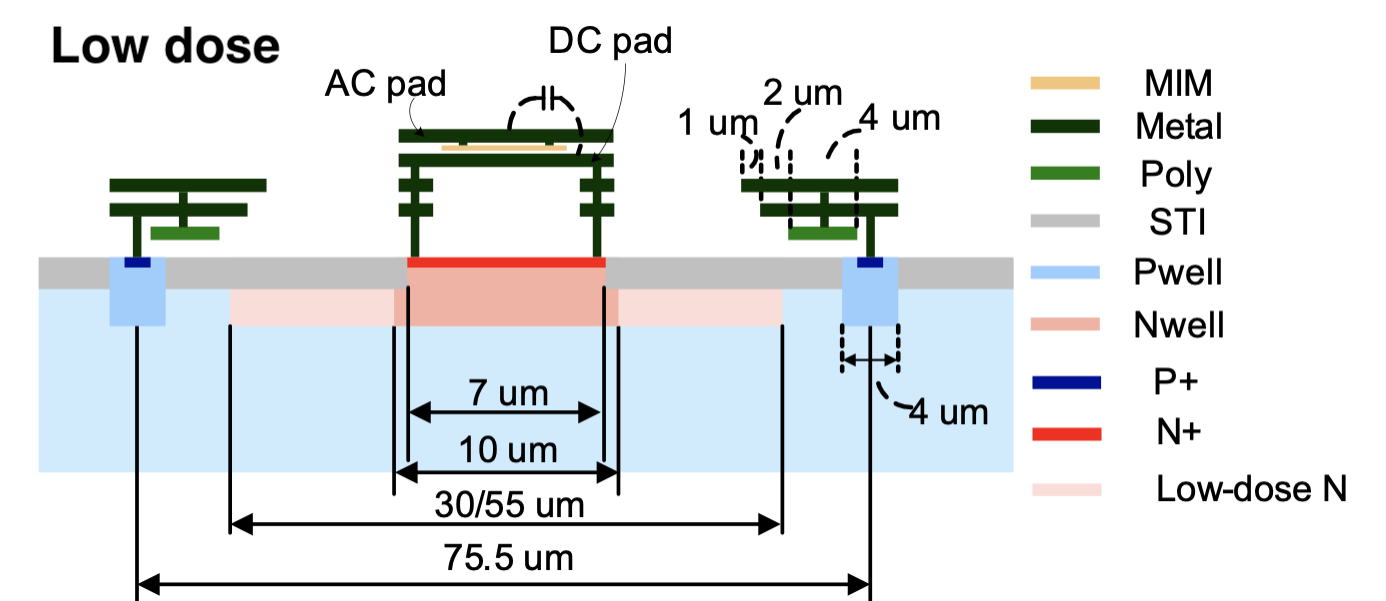}
    \caption{ \small Low dose strip design sketch. It features two low-dose n-wells widths, 30 and 55~$\upmu\mathrm{m}$.
    \label{fig:low dose}}
\end{figure}

As shown in figure \ref{fig:long_picture}, each die houses two strip detectors, each one with 40 strips. There is a \emph{regular design} (see figure \ref{fig:regular}) which is similar to the ATLAS experiment phase 2 upgrade strip design, and a \emph{low dose design} (see figure \ref{fig:low dose}) which has some features of the CMOS technology steps such as including a MIM (metal-insulator-metal) capacitor and an n-well implant. The n-well implant of the \emph{low dose design} has two different implant widths, 20 strips have a \SI{30}{\micro\m} implant width, and 20 strips have a \SI{55}{\micro\m} n-well implant width. %For the radioactive source measurements the results are separated by the \SI{30}{\micro\m} and the \SI{55}{\micro\m} implants.% (and for short it is written as LD30 and LD55 respectively).

A first sensor batch underwent a backside processing that was not optimized and had early break down once the sensors were reaching full depletion, shown in Ref. \cite{a}. For the next batches, the backside processing was improved and those sensors have stable current up to \SI{300}{\V}, they reach full depletion at $\sim$\SI{35}{\V}, with excellent results and not showing any effect of the stitching. Those sensors were irradiated with neutrons, tested in beam lines, with transient current technique and charge collection lab set-ups\cite{b,c,d,e,f,g}. Simulations about possible stitching scenarios show that stitching should not affect the performance of  the implant of the strip \cite{simulations}.

%The detectors have two lengths, \emph{short} are \SI{2.1}{\cm} and \emph{long} are \SI{4.1}{\cm} long (figure \ref{fig:long_picture} shows a picture of the long sensor). They have 3 or 5 stitching lines respectively. 

%it was shown that passive strip detectors work very well although those results were done with a first fabrication that the backplane was unsuccessfully processed and without studying the performance after irradiation. This paper shows the results with the same detectors but with an improved backplane processing which decreased the leakage current and the break down voltage is beyond \SI{300}{\V}.

Here we will show the results of passive CMOS strip detectors irradiated at CERN\cite{irrad} in the IRRAD facility with \SI{24}{\giga\electronvolt} protons at \SI{5e14}{n_{eq}/\cm^2} and \SI{1e15}{n_{eq}/\cm^2} fluences to study the effects of proton irradiation. The irradiation collimator is \SI{1}{\cm^2} big, therefore the sensors were tilted during irradiation in order to receive the particle beam along the full area. The irradiations were carried out at room temperature. 

This paper shows the electrical characterization and charge collection measurements with a $^{90}$Sr radioactive source for the proton irradiated passive CMOS strip detectors, with the aim to investigate the effects of proton irradiation. 

%%%%%%%%%%%%%%%%%%%%%%%%%%%%%%%%%%%%%%%%%%%%%%%%%%%%%%%%

\section{Electrical characterisation}

\begin{figure}[ht]
    \centering
    \includegraphics[width=0.95\linewidth]{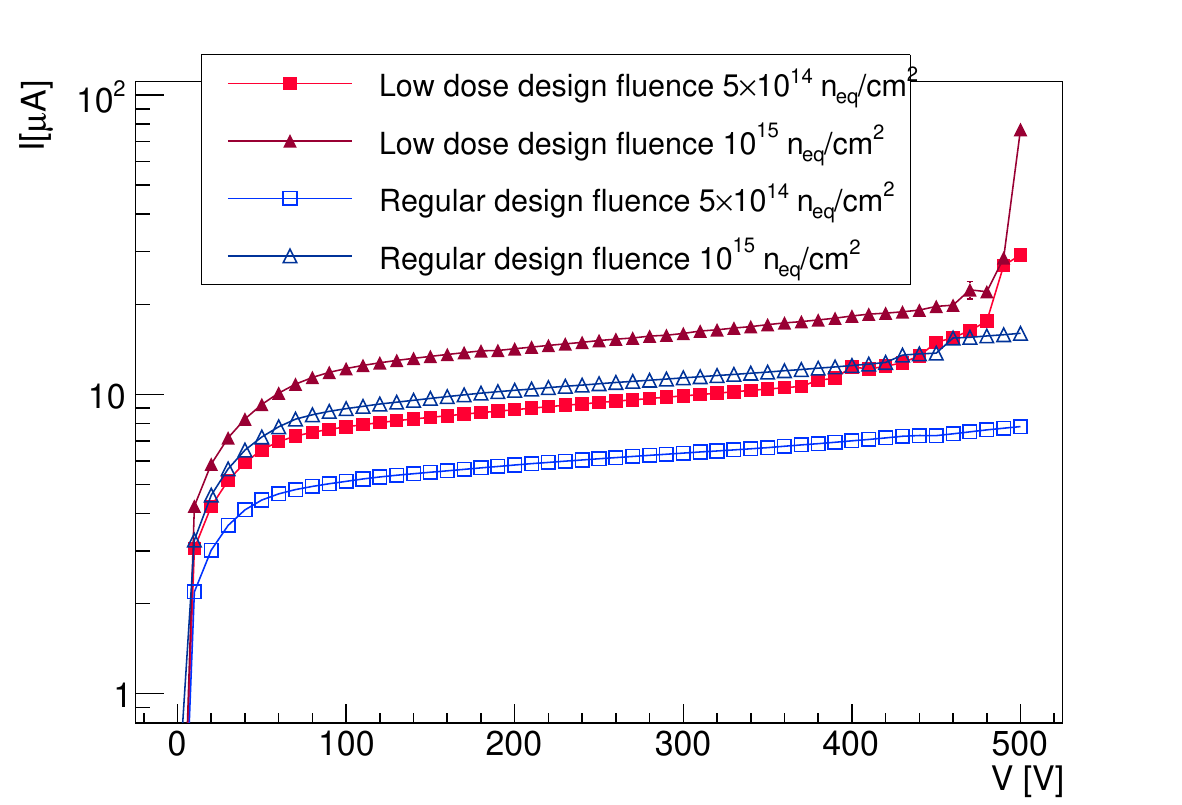}
    \caption{ \small Current-voltage characteristics of the sensors after irradiation at \SI{-20}{\degreeCelsius}.
    \label{fig:iv_no_annealed}}
\end{figure}

% canviar plots: treure el de mínima fluencia
% treure titols
Current voltage characteristics of the reverse bias were measured with a wafer prober MPI-TS3500-SE, connected to a Keithley 2410 high voltage power supply. High voltage was applied to the sensors between the bias pad and the backplane. Figure \ref{fig:iv_no_annealed} shows the current voltage characteristics of the sensors for the two different fluences without annealing. Measurements were taken at \SI{-20}{\degreeCelsius} and show the average of five consecutive measurements per voltage step. The detectors used for this study are the \SI{4.1}{\cm} long sensors.

%For an unirradiated sensor, the leakage current was $\sim$ \SI{30}{\nano\ampere}

\begin{figure}[ht]
    \centering
    \includegraphics[width=0.95\linewidth]{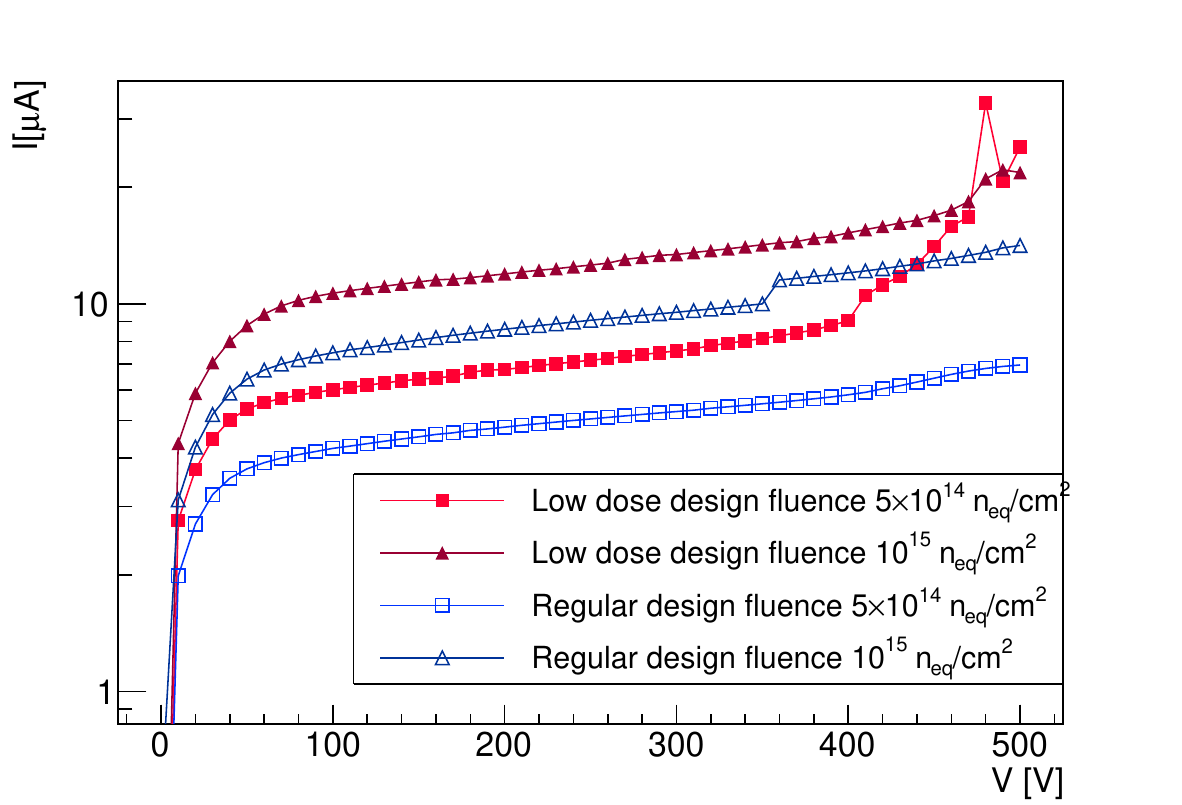}
    \caption{ \small Current-voltage characteristics of the sensors after irradiation at \SI{-20}{\degreeCelsius}, with an annealing of \SI{80}{\minute} at \SI{60}{\degreeCelsius}.
    \label{fig:iv_annealed}}
\end{figure}

Afterwards, the sensors were annealed for \SI{80}{\minute} at \SI{60}{\degreeCelsius}. Figure \ref{fig:iv_annealed} shows the current voltage characteristics after annealing. As expected, after the beneficial annealing the leakage current decreases. The break down voltage is not yet reach at \SI{500}{\volt}.

\begin{figure}[ht]
    \centering
    \includegraphics[width=0.95\linewidth]{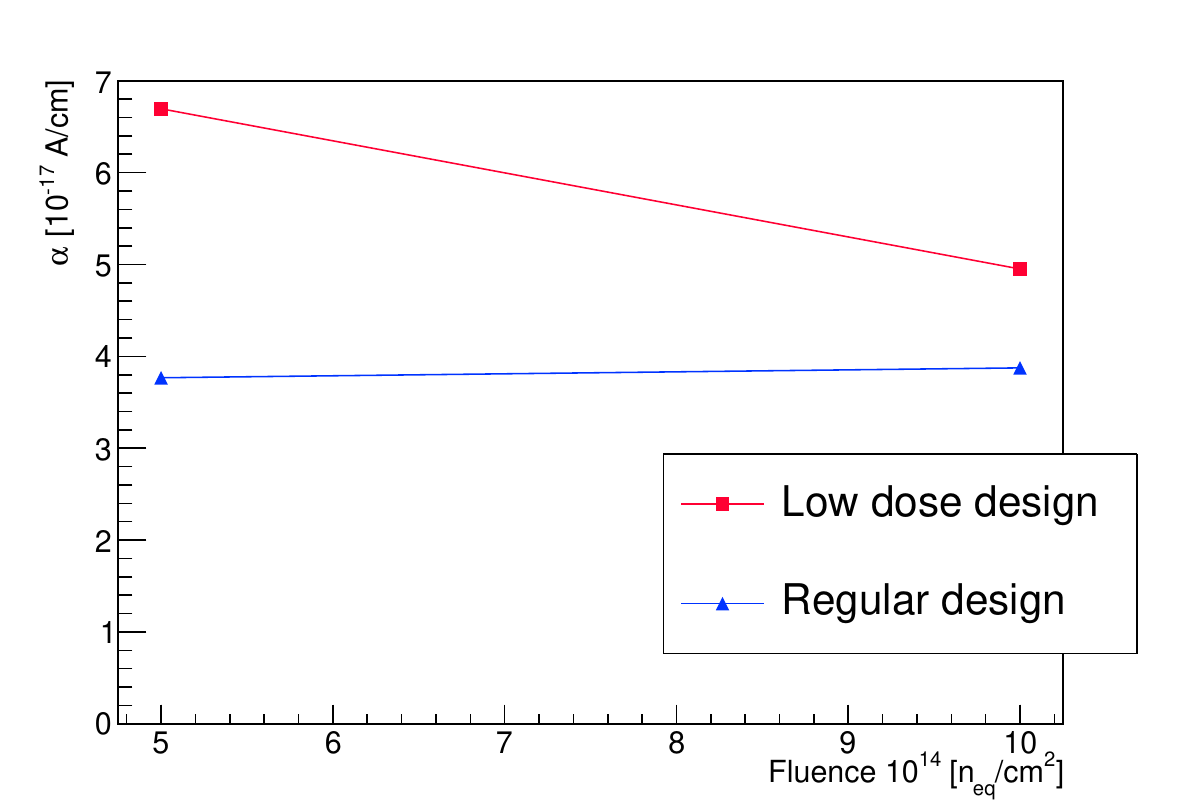}
    \caption{ \small Current-related damage rate at \SI{400}{\V} of the passive CMOS strip detectors for different fluences, with an annealing of \SI{80}{\minute} at \SI{60}{\degreeCelsius}.
    \label{fig:alpha}}
\end{figure}

From the leakage current the current-related damage rate ($\alpha$) can be calculated with the expression\cite{alpha}:

$$ \Delta I_L=\alpha\phi_{eq}V$$

where $\Delta I_L$ is the difference between leakage currents at \SI{20}{\degreeCelsius} due to irradiation, $\phi_{eq}$ is the neutron equivalent fluence and $V$ is the volume of the sensor. The temperature scaling from \SI{-20}{\degreeCelsius} to \SI{20}{\degreeCelsius} is done with the method described in Ref. \cite{chilingarov}. Figure \ref{fig:alpha} shows the current-related damage rate at \SI{400}{\V} calculated from the leakage current of figure \ref{fig:iv_annealed} (after annealing). Low dose design shows a higher current-related damage rate than the regular design since they show higher leakage current.

\begin{figure}[ht]
    \centering
    \includegraphics[width=0.95\linewidth]{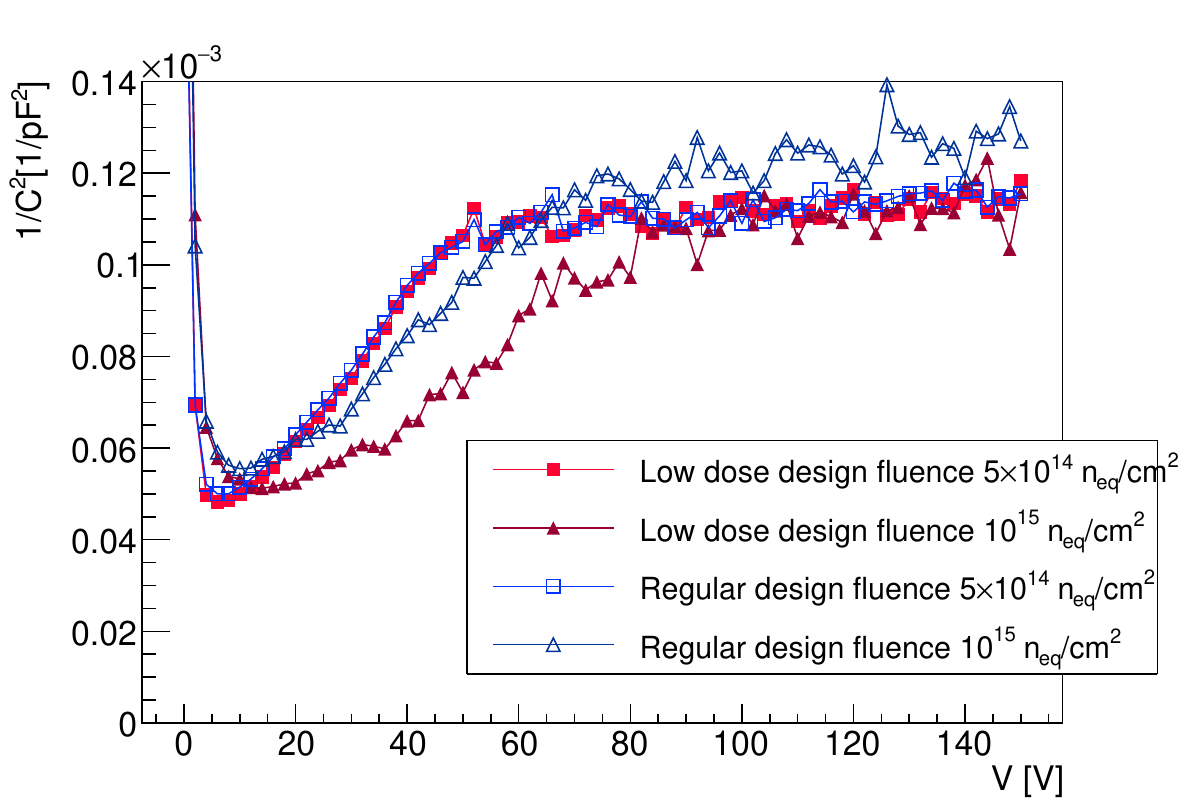}
    \caption{ \small Capacitance voltage characteristics of the sensors after irradiation at \SI{-20}{\degreeCelsius}, with an annealing of \SI{80}{\minute} at \SI{60}{\degreeCelsius}.
    \label{fig:cv_annealed}}
\end{figure}

Moreover, to investigate the full depletion voltage after irradiation, a capacitance voltage measurement was conducted with a HP4284A LCR-meter. The measurements were taken at a \SI{1}{\kilo\hertz} frequency and at \SI{-20}{\degreeCelsius}.
Figure \ref{fig:cv_annealed} shows the capacitance voltage characteristics for the sensors after annealing. The capacitance reaches a plateau after $\sim$\SI{50}{\volt} for the fluence of \SI{5e14}{n_{eq}\per\cm^2}, while for the \SI{1e15}{n_{eq}\per\cm^2} irradiated sensors, the regular design reach a plateau at $\sim$\SI{70}{\volt} and the low dose design at $\sim$\SI{80}{\volt}.

From the depletion voltage, the effective doping can be estimated using the relation:

$$N_{eff}=\frac{2\epsilon_0\epsilon_{Si}V_{FD}}{q\cdot d^2}$$

\begin{figure}[ht]
    \centering
    \includegraphics[width=0.95\linewidth]{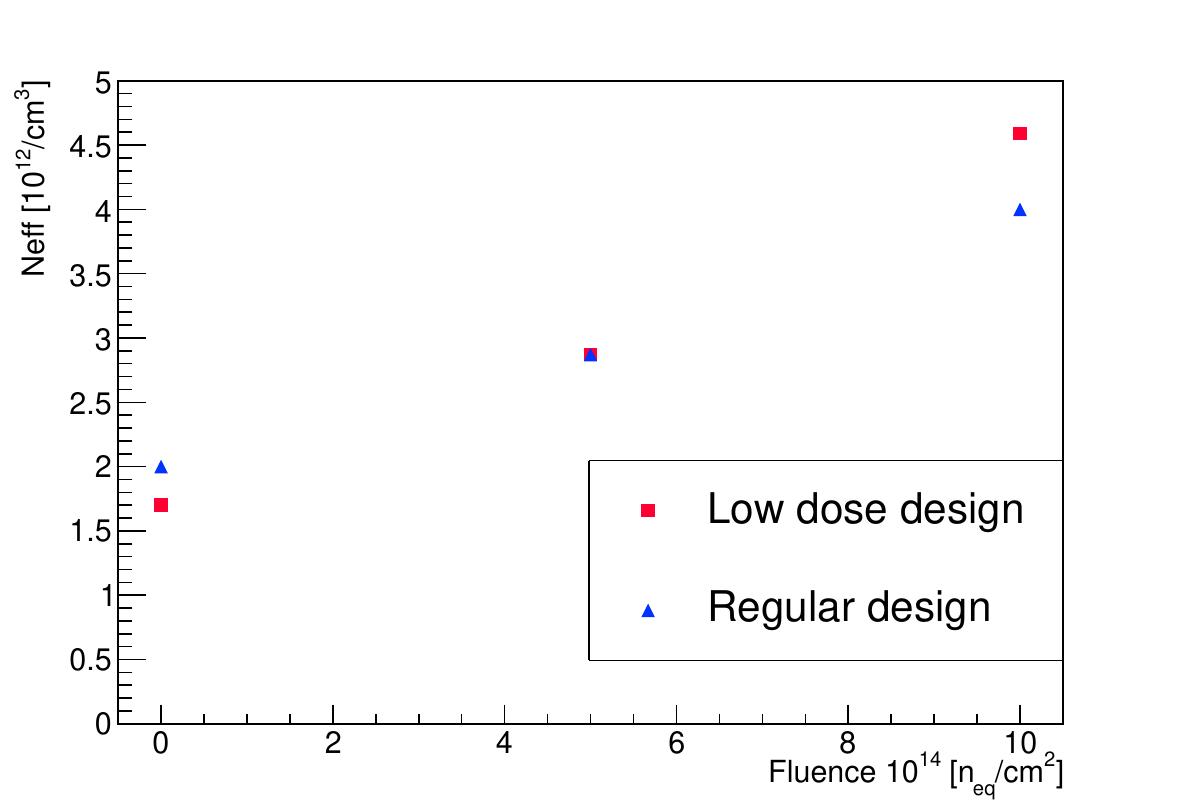}
    \caption{ \small Effective doping of the sensors at different fluences.
    \label{fig:neff}}
\end{figure}

Where $\epsilon_0$ is the vacuum permittivity, $\epsilon_{Si}$ is the silicon permittivity, $V_{FD}$ is the full depletion voltage, $q$ is the electron charge and $d$ is the sensor thickness. For an unirradiated sample, the $V_{FD}$ is \SI{30}{\V} for low dose design and \SI{35}{\V} for the regular design, that gives an effective doping of \SI{1.7e12}{\cm^{-3}} and \SI{2e12}{\cm^{-3}} respectively. Figure \ref{fig:neff} shows the effective doping with the fluence. Although the bulk doping should be the same for regular and low dose designs, the effective doping shows a dependence on the strip design, inverting the behaviour of the low dose design: for unirradiated sensors low dose design show lower effective doping than the regular design, while for higher fluences the low dose design shows higher effective doping than the regular design.

%Considering that both designs deplete at \SI{50}{\V} when irradiated with a fluence of \SI{5e14}{n_{eq}\per\cm^2} that corresponds to an effective doping of \SI{2.87e12}{\cm^{-3}}. After the sensors being irradiated with a fluence of \SI{1e15}{n_{eq}\per\cm^2} they show a full depletion voltage of \SI{70}{\V} for the regular design and \SI{80}{\V} for the low dose design, corresponding to an effective doping of \SI{4e12}{\cm^{-3}} and \SI{4.59}{\cm^{-3}} respectively. 

\section{Charge collection measurements}

The two irradiated sensors were then bonded to an ALiBaVa Daughter board\cite{z}. The daughter board is then connected to the ALiBaVa mother board which reads out the bonded channels and the data is acquired with a computer. A scintillator was located under the sensor to externally trigger the signal. All the measurements were taken at \SI{-20}{\degreeCelsius}. On top of the sensor, a $^{90}$Sr beta source was positioned that was simultaneously exposing the sensor and the scintillator to electrons. %The sensors were annealed during \SI{80}{\minute} at \SI{60}{\degreeCelsius}.

\begin{figure}[ht]
    \centering
    \includegraphics[width=0.95\linewidth]{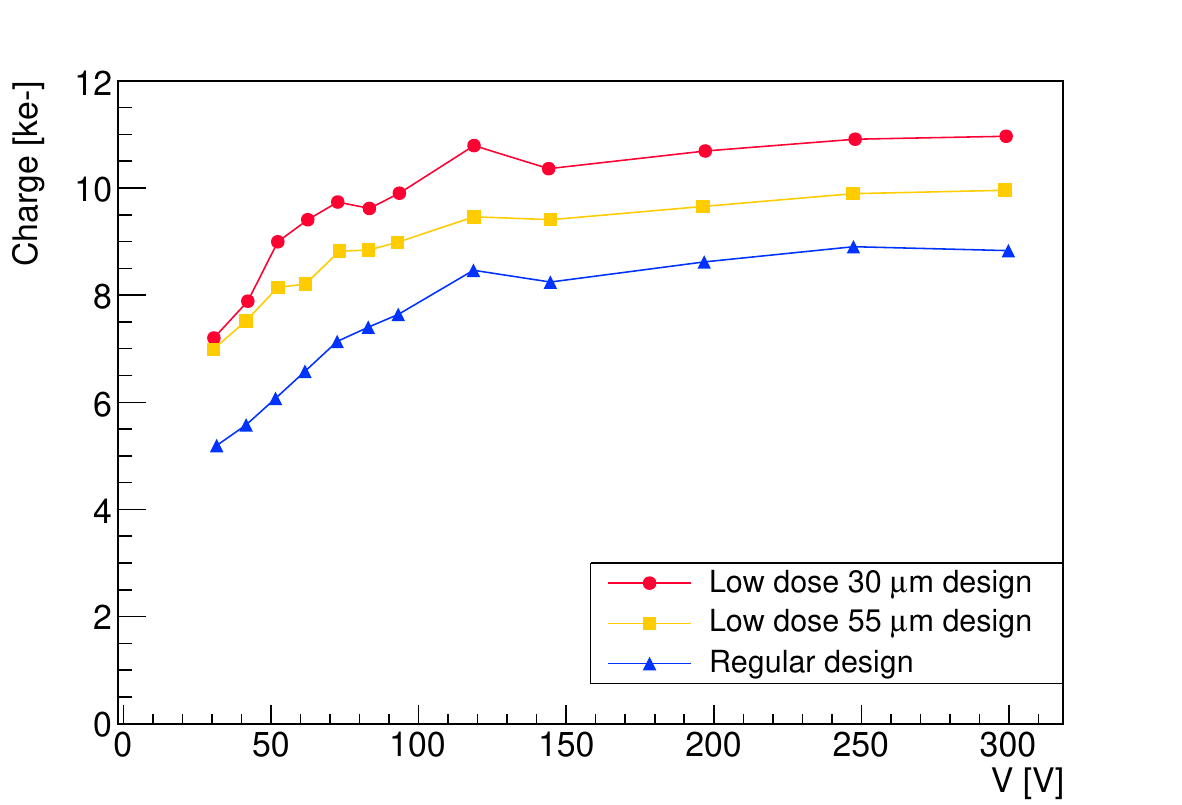}
    \caption{ \small Collected charge measurement at  \SI{-20}{\degreeCelsius} from a sensor irradiated with a fluence of \SI{5e14}{n_{eq}\per\cm^2}, after an annealing of \SI{80}{\minute} at \SI{60}{\degreeCelsius}.
        \label{fig:cc_5e14}}
\end{figure}

\begin{figure}[ht]
    \centering
    \includegraphics[width=0.95\linewidth]{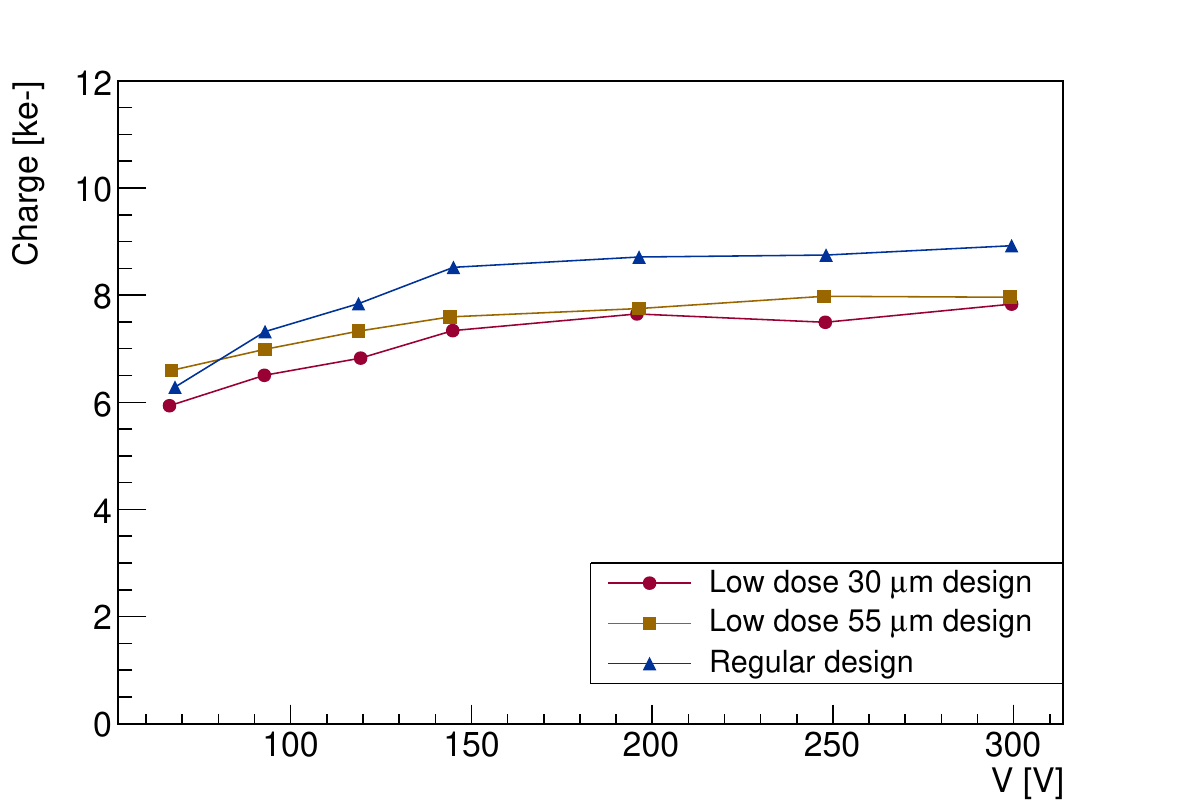}
    \caption{ \small Collected charge measurement at  \SI{-20}{\degreeCelsius} from a sensor irradiated with a fluence of \SI{1e15}{n_{eq}\per\cm^2}, after an annealing of \SI{80}{\minute} at \SI{60}{\degreeCelsius}.
    \label{fig:cc_1e15}}
\end{figure}

Figure  \ref{fig:cc_5e14} and figure  \ref{fig:cc_1e15} show the collected charge of the detector after being irradiated with \SI{5e14}{n_{eq}\per\cm^2} and \SI{1e15}{n_{eq}\per\cm^2} respectively and annealed for \SI{80}{\minute} at \SI{60}{\degreeCelsius}. For this measurement, low dose \SI{30}{\micro\m} and low dose \SI{55}{\micro\m} are separated since individual strips can be masked. The chips were previously calibrated with another sensor, therefore the charge is shown in electrons instead of Analogue to Digital Converter units.

Figure \ref{fig:cc_5e14} (sensor irradiated at \SI{5e14}{n_{eq}\per\cm^2} fluence) shows higher collected charge for the low dose designs, almost equal to unirradiated values ($\sim$\SI{11000}{electrons}), while the regular design has a noticeable decrease of collected charge. On the other hand, the sensors irradiated with \SI{1e15}{n_{eq}\per\cm^2} shown in figure \ref{fig:cc_1e15} have the opposite behaviour. The alpha parameter decreases for the low dose design, while staying nearly constant for the regular design. This behaviour cannot be associated to the bulk damage since that should be the same for all the designs but to differences from the strip design. Besides the differences between designs, the results show very good charge collection for passive CMOS strip detectors after proton irradiation and do not show any stitching effect. %The collected charge for the $^{90}$Sr beta source shows a decrease of the collected charge for the low dose designs while the regular design collected charge remains very similar for both fluences.  

\section{Summary and Conclusion}

Passive CMOS strip detectors do not show any impact from stitching in their performance as shown here and in previous results. The irradiation with protons shows some differences between the two designs used, low dose and regular, decreasing the performance of low dose design for higher fluences.  

The successful performance of the passive CMOS strip detectors and the fact that the stitching does not show any effect for particle detection opens the door for the CMOS foundries to fabricate strip sensors. Future experiments may use stitched strip sensors since it is not affecting the efficiency. Besides that, future strip productions are investigating to include active parts, with electronics embedded in the silicon, therefore producing the first Monolithic Active Strip Sensors (MASS) and avoid future hybridisation steps for strips.

\section{Acknowledgements}

The authors acknowledge the usage of the wafer prober at the TU Dortmund. Funded by the Deutsche Forschungsgemeinschaft (DFG, German Research Foundation)- 450639102.

{}

\end{document}